\documentclass[12pt]{iopart}
\usepackage{graphicx}
\usepackage{subcaption}
\usepackage{float} 
\begin{document}

\title[Correlation of the L-mode density limit with edge collisionality]{Correlation of the L-mode density limit with edge collisionality}

\author{A D Maris$^1$, C Rea$^1$, A Pau$^2$, W Hu$^3$,  B Xiao$^3$, R Granetz$^1$, E Marmar$^1$, the EUROfusion Tokamak Exploitation team*, the Alcator C-Mod team, the ASDEX Upgrade team**, the DIII-D team, the EAST team, and the TCV team***}

\address{$^1$ Plasma Science and Fusion Center, Massachusetts Institute of Technology, Cambridge, MA 02139, USA}
\address{$^2$École Polytechnique Fédérale de Lausanne (EPFL), Swiss Plasma Center (SPC), CH-1015 Lausanne, Switzerland}
\address{$^3$Institute of Plasma Physics, Chinese Academy of Sciences, Hefei 230031, CN}
 \address{* See the author list of E. Joffrin et al Nucl. Fusion 2024}
 \address{** See the author list of U. Stroth et al 2022 Nucl. Fusion 62, 042006}
 \address{*** See author list of H. Reimerdes et al 2022 Nucl. Fusion 62 042018}
\ead{maris@mit.edu}
\vspace{10pt}
\begin{indented}
\item[]December 2024
\end{indented}

\begin{abstract}
The ``density limit'' is one of the fundamental bounds on tokamak operating space, and is commonly estimated via the empirical Greenwald scaling. This limit has garnered renewed interest in recent years as it has become clear that ITER and many tokamak pilot plant concepts must operate near or above the Greenwald limit to achieve their objectives. Evidence has also grown that the Greenwald scaling - in its remarkable simplicity - may not capture the full complexity of the density limit. In this study, we assemble a multi-machine database to quantify the effectiveness of the Greenwald limit as a predictor of the L-mode density limit and compare it with data-driven approaches. We find that a boundary in the plasma edge involving dimensionless collisionality and pressure, $\nu_{*\rm, edge}^{\rm limit} = 3.5 \beta_{T,{\rm edge}}^{-0.40}$, achieves significantly higher accuracy (false positive rate of 2.3\%  at a true positive rate of 95\%) of predicting density limit disruptions than the Greenwald limit (false positive rate of 13.4\% at a true positive rate of 95\%) across a multi-machine dataset including  metal- and carbon-wall tokamaks (AUG, C-Mod, DIII-D, and TCV). This two-parameter boundary succeeds at predicting L-mode density limits by robustly identifying the radiative state preceding the terminal MHD instability. This boundary can be applied for density limit avoidance in current devices and in ITER, where it can be measured and responded to in real time.
\end{abstract}

%
\vspace{2pc}

\submitto{\NF}

\maketitle
 

\section{Introduction} \label{sec:intro}

Plasma electron density ($n_e$) is a critical lever for fusion performance in tokamaks. High density is necessary for many burning plasma tokamak concepts to maximize fusion triple product $nT\tau_E$ \cite{mcnally1977ignition}, enhance bootstrap current drive (via steeper density gradients) \cite{buttery2021advanced}, and enable divertor detachment \cite{krasheninnikov2016divertor}. There has long been an interest in developing scaling laws to describe the highest achieveable density in tokamaks \cite{murakami1976some,fielding1977high,greenwald2002density}. Today, the most widely utilized empirical density limit scaling is the ``Greenwald limit'' \cite{greenwald1988new}, expressed as
\begin{equation} \label{eq:greenwald}
    \frac{\bar{n} }{n_G} = 1,
\end{equation}
where $\bar{n}$ is the central line-averaged electron density in units of $10^{20}$ m$^{-3}$, $n_G \equiv I_p/\pi a^2$ is the ``Greenwald density,'' $I_p$ is the plasma current in MA, and $a$ is the minor radius in meters. Operating near or above this limit correlates with confinement regime transitions when the plasma is in the ``high'' confinement mode (H-mode) and disruptions when the plasma is in the ``low'' confinement mode (L-mode). Nevertheless, to maximize fusion power, burning plasma experiments such as ITER \cite{shimada2007overview} and fusion power plant (FPP) concepts (such as EU-DEMO \cite{giruzzi2015modelling}, the compact advanced tokamak \cite{buttery2021advanced}, and ARC \cite{sorbom2015arc}) are designed to operate near or above the Greenwald limit. Of course, by choosing to operate near an instability limit, future devices run the risk of harmful transients, such as H-to-L back-transitions and disruptions. Even infrequent unmitigated disruptions - such as once a month - could render tokamak power plants uneconomical given the long timescales needed for repairs \cite{maris2023impact}. Therefore, tokamak power plants require large safety margins and/or extremely effective control solutions for the density limit and other instabilities.\\

\begin{figure} [!htb]
    \centering
    \includegraphics[width=\linewidth]{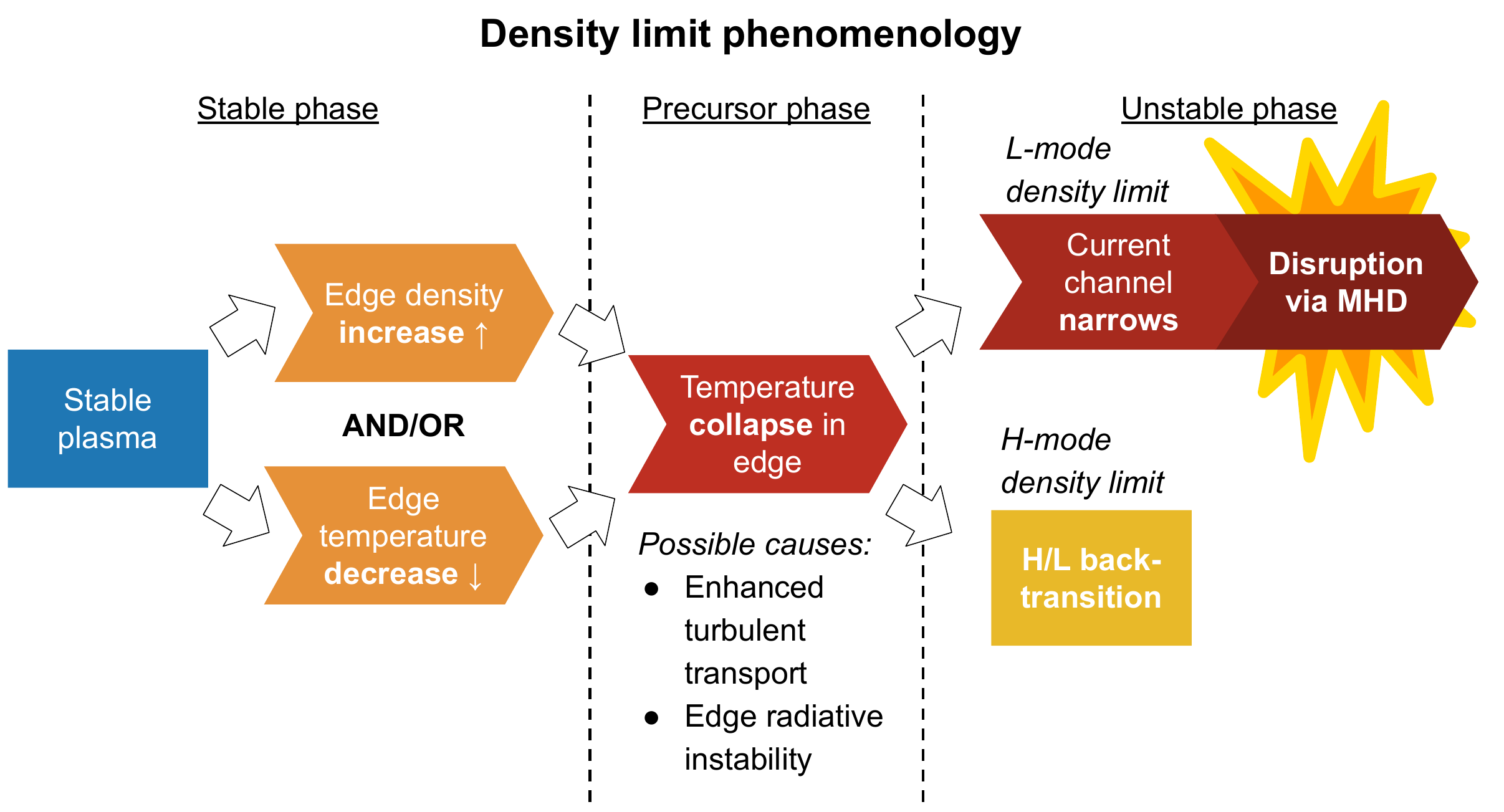}
    \caption{The characteristic chain of events for density limits in tokamaks. If the discharge is in H-mode, it suffers an H-to-L back-transition; this is referred to as an ``H-mode density limit.'' If it is in L-mode, the current channel shrinks, ultimately resulting in an ``L-mode density limit'' disruption.}
    \label{fig:dl_chain}
\end{figure}

While a complete, first-principles treatment of the density limit remains elusive, theory and experiment have clarified the characteristic dynamics, summarized in Fig. \ref{fig:dl_chain}. The path to the density limit begins with edge density increasing and/or edge temperature decreasing \cite{labombard2001particle}. Past a certain threshold, a thermal instability occurs at the plasma edge, causing a collapse of the edge temperature. If the plasma is operating in H-mode, it experiences an H-to-L back-transition, referred to as an ``H-mode density limit'' (HDL). In L-mode, this temperature collapse is associated with the formation of an X-point radiator or a MARFE - a toroidally symmetric ring of cool, dense plasma on the high-field side \cite{lipschultz1984marfe}. The HDL is not a disruptive instability itself, but can be followed by the ``L-mode density limit'' (LDL). The LDL occurs when the edge cooling causes the plasma current to concentrate in a peaked current profile \cite{shi2017first,li2023local}. When current channel is sufficiently narrow, it loses MHD stability, causing a disruption.

Theories attempting to explain the thermal instability tend to come in two flavors: a radiative instability in the edge \cite{stroth2022model,zanca2019power} or enhanced turbulent transport in the edge \cite{rogers1998phase,giacomin2022first,eich2021separatrix,singh2022zonal}. It has been shown that many of these models share qualitative similarities to each other \cite{manz2023power}.

This sharpening picture of the density limit suggests that burning plasmas may be able to exceed the Greenwald limit for two reasons. Because the density limit depends on the edge of the plasma, experiments with density peaking have achieved $\bar{n}/n_G > 1$ while maintaining $n_{\rm edge}/n_G < 1$ \cite{gibson1990fusion,kamada1991study,stabler1992density,osborne2001gas,pucella2017analytical}. It is expected that burning plasmas will naturally exhibit density peaking due to their low collisionality  \cite{angioni2007scaling}. Additionally, several studies following Ref. \cite{greenwald1988new} have observed a modest input power scaling of $P^{0.2-0.6}$ augmenting the Greenwald limit \cite{gibson1990fusion,giacomin2022first,manz2023power,kamada1991study,stabler1992density,rapp1999density,huber2013impact,bernert2014h}. This power dependence is understood to be due to higher input power raising the temperature at the edge and holding off the onset of the temperature instability.

At the same time, we will show in this paper that these two observations alone are not sufficient to achieve the extremely high disruption prediction accuracy required for ITER \cite{lehnen_plasma_2018}. The Greenwald limit was not derived with disruption prediction in mind, and so it is perhaps not surprising that there is room for improvement for a density limit disruption forecaster. It is notable however, that we must go beyond just applying a power scaling and using the edge density, we must instead combine edge temperature and density to predict LDLs with high accuracy.\\

In this paper, we assemble and study a multi-machine database of LDL events from ASDEX UpGrade (AUG), Alcator C-Mod, DIII-D, EAST, and TCV. We apply a variety of techniques to predict the onset of the instability, thus finding:
\begin{itemize}
    \item[1)] The Greenwald limit does not universally predict the onset of L-mode density limit events (false positive rate of $>\,$13\% for a true positive rate of 95\%).
    \item[2)] Data-driven models trained to predict the radiative precursor phase achieve significantly improved L-mode density limit prediction accuracy (false positive rate of $<\,$3\% for a true positive rate of 95\%). 
    \item[3)] In particular, a simple stability boundary in terms of the effective collisionality and $\beta_T$ in the plasma edge, $\nu_{*\rm, edge}^{\rm limit} = 3.5 \beta_{T,\rm edge}^{-0.40}$, is a highly reliable proximity-to-density-limit metric (false positive rate of 2.3\% for a true positive rate of 95\%).  
\end{itemize}

The paper is organized as follows: Section \ref{sec:methods} describes the methods used for the dataset assembly and the data-driven analysis, Section \ref{sec:results} describes the prediction performance of various models on an unseen test set, Section \ref{sec:discuss} discusses the relation to existing density limit observations and considers example discharges from the database, and Section \ref{sec:conc} summarizes the findings of this study and outlines future work.

\section{Methods} \label{sec:methods}

\subsection{Dataset and labeling} \label{subsec:dataset}

The dataset utilized for this study is composed of discharges from five tokamaks: AUG, C-Mod, DIII-D, EAST, and TCV. The C-Mod and DIII-D database of LDLs are newly collated for this study based on data fetching workflows utilized in Ref. \cite{rea2018disruption, montes2019machine}. The LDL shots from AUG and TCV in this study appeared in Ref. \cite{giacomin2022first}, and those from EAST appeared in Ref. \cite{hu2023prediction}. The number of discharges and samples from each device is shown in Table \ref{tab:num_shots}, and the global parameters of these devices can be found in \ref{sec:macro_params_corr_matrix}, Table \ref{tab:eng_sum}. The reader should note the significant variation in the number discharges available for each device due to different data availability, frequency of density limit experiments, and lifetimes of the machines.

\begin{table}
\caption{ Number of unique discharges and time steps in the database assembled for this study, divided into discharges that feature an L-mode density limit (LDL) and those that do not (stable).}
\label{tab:num_shots}
\begin{indented}
\item[]\begin{tabular}{@{}lllll}
\br
Device & LDL discharges & LDL time steps & Stable discharges & Stable time steps \\
\mr
AUG & 33 & 1,106 & 8 & 16,231 \\
C-Mod & 92 & 3,819 & 2,429 & 275,492 \\
DIII-D & 42 & 2,225 & 1,073 & 367,171 \\
EAST & 13 & 1,498 & 669 & 683,750 \\
TCV & 32 & 1,511 & 74 & 11,004 \\
\br
Total & 212 & 10,159 & 4,253 & 1,353,648 \\

\end{tabular}
\end{indented}
\end{table}

\begin{figure}[!htbp]
    \centering
    \includegraphics[width=\linewidth]{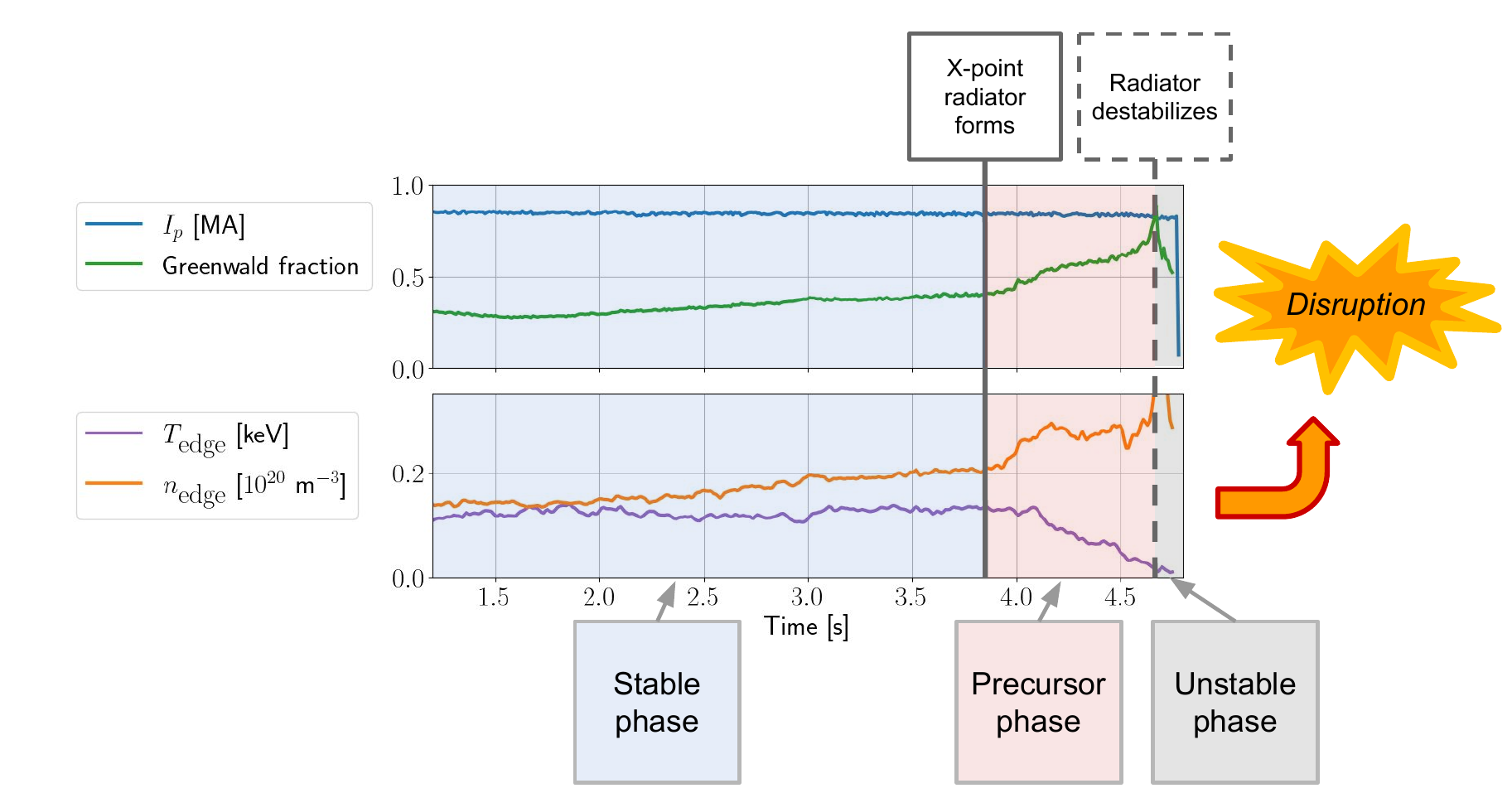}
    \caption{Labeling the L-mode density limit for an example discharge, DIII-D \#191794. The LDL precursor phase is the time period between when the X-point radiator forms and when the radiator becomes unstable and moves toward the plasma core. We ignore the window in time after the radiator is destabilized. Here, ``edge'' density and temperature are defined as the average measurement of the quantity between $\rho =$ 0.85 and 0.95.}
    \label{fig:label}
\end{figure}

Density limit discharges were manually labeled by the authors using the pattern observed across all devices: an increase in density and/or decrease in edge temperature, followed by the formation of a radiator or MARFE near the X-point, which eventually destabilizes and moves towards the core, resulting in a disruption. For this study, the LDL precursor start time was labeled manually as the time of the X-point radiator formation when such measurements are available (AUG, C-Mod, DIII-D, and TCV) and a fixed 100 ms window before the MARFE destabilizes otherwise (EAST). In AUG and TCV, the DEFUSE tool \cite{pau2023modern} automatically tags candidate events which are subsequently manually validated by an expert. For DIII-D, the formation time is determined by manual inspection of individual bolometer chords and 2D tomographic reconstructions of the poloidal radiation cross-section. For C-Mod, this is done via inspection of an H-alpha chord and visible camera images. The LDL precursor end time was manually labeled to occur as the radiation front moves inward toward the core of the plasma before the disruption. An example of this labeling is schematically represented in Figure \ref{fig:label}.

To isolate density limit dynamics from other instabilities, LDL candidates were excluded from the database if 1) the operators noted major impurity injections, 2) there was significant MHD activity prior to the formation of the X-point radiator, 3) or the disruption was immediately proceeded by a sudden shutoff or failure of input power.

Non-disruptive discharges, also referred to here as ``stable'' discharges, are uniformly sampled from the set of discharges in each device that did not result in a disruption during flattop and did not experience control errors. Stable discharges that experienced minor disruptions were also excluded. A correlation matrix for the dataset is reported in \ref{sec:macro_params_corr_matrix}.

\subsection{Feature set} \label{subsec:features}

Table \ref{tab:feature_list} lists the signals, which we refer to as ``features'', used in this analysis. Edge density and temperature in this study are defined as the Thomson Scattering measurements averaged between a normalized radius ($\rho$) of 0.85 and 0.95, as was used for experimental validation of an edge density limit in Ref. \cite{giacomin2022first}. A simple fitting procedure was used to determine the profiles of AUG and TCV, while linear interpolation was used for C-Mod and DIII-D. Averaging over this relatively large edge region reduces the impact of measurement noise. Normalized radius is defined in terms of the square root of the normalized toroidal flux for DIII-D data and the square root of the normalized poloidal flux for AUG, TCV, and C-Mod data. All signals are resampled onto a 10ms timebase for consistency.

\begin{table}
\caption{Features in the dataset, as well as ``features sets'' used in analysis: ``global'' features, ``edge'' features, and dimensionless features.}
\label{tab:feature_list}
\begin{indented}
\item[]\begin{tabular}{@{}llccc}
\br
Symbol & Definition &  \shortstack{``Global'' \\ features} & \shortstack{``Edge'' \\ features} & \shortstack{Dimensionless \\ features} \\
\mr
$\bar{n}$ & e\textsuperscript{-} density, line avg. & X &  & \\
$n_{\rm edge}$ & e\textsuperscript{-} density, edge & & X  & \\
$P_{\rm in}$ & Input power & X &  & \\
$T_{\rm edge}$ & e\textsuperscript{-} temperature, edge & & X   &\\
$I_p$ & Plasma current & X & X  & \\
$a$ & Minor radius & X & X &  \\
$q_{95}$ & Safety factor &  & X & X \\
$\nu_{*\rm, edge} $ & Collisionality, edge &  &  & X \\
$\beta_{T,\rm edge}$ & Toroidal $\beta$, edge &  &  & X \\
$\rho_{*,\rm edge} $ & Norm. gyroradius, edge &  &  & X \\
\br
\end{tabular}
\end{indented}
\end{table}

As shown in Table \ref{tab:feature_list}, we conduct our analyses using three distinct sets of features: ``global'' features, ``edge'' features, and dimensionless features. The global features are macroscopic plasma parameters that are relatively easy to measure (ex. $\bar{n}$, $I_p$) and typically utilized in density limit scalings. The edge features are similar, but with the line-averaged density and input power replaced with the edge density and temperature, respectively. These latter two parameters are expected to be more predictive of the density limit because they are local to the edge region where the density limit is thought to be triggered. We also add the edge safety factor, $q_{95}$, which may capture additional information related to the connection length at the edge of the plasma. Because of noise in the measurement of edge density and temperature, a Butterworth filter is applied to these signals with a critical frequency of 8 Hz and 6 Hz, respectively. For the sake of cross-device consistency, the same filter is applied to all devices.

Due to the strong correlation in our dataset between minor radius $a$, major radius $R_0$, and the toroidal magnetic field $B_T$, our primary results will only include the minor radius as in the Greenwald scaling. We seek to avoid multicollinearity in training models because it can mask true variable interactions. For the same reason, we do not train models with both line-average density $\bar{n}$ and edge density $n_{\rm edge}$. Additionally, we have measurements for inverse aspect ratio $\epsilon$, elongation $\kappa$, and triangularity $\delta$ across our database, but we exclude them from the analysis as there is too little variation among these parameters to be of use in this study.

We also analyze a set of dimensionless variables generally following the definitions used in \cite{verdoolaege2021updated}, but with ion density and temperature replaced with the electron value. The dimensionless variables we use include $q_{95}$ and the following three variables:
\begin{equation}
    \nu_{\rm *,edge} \equiv \frac{ \nu_{ii} q_{\rm cyl} R_0}{ v_{ti} \epsilon^{3/2} } \approx  \frac{e^4 \ln (\Lambda)}{2 \pi \epsilon_0^2} \frac{n_{\rm edge}}{ T_{\rm edge}^2 }  \frac{ q_{\rm cyl} R_0}{ \epsilon^{3/2} }, 
\end{equation}
\begin{equation}
    \rho_{*,\rm edge} \equiv \frac{\rho_i}{a} \approx \frac{ \sqrt{m_i T_{\rm edge}}, }{e B_T a},
\end{equation}
\begin{equation}
    \beta_{T,\rm edge} \equiv \frac{ 2 n_{\rm edge} T_{\rm edge} }{B_T^2/(2\mu_0)},
\end{equation}
where $\nu_{ii}$ is the ion-ion collision frequency, $q_{\rm cyl} \equiv \frac{2 \pi}{\mu_0} \frac{B_T a^2}{I_p R_0} (\frac{1 + \kappa^2 }{2})$ is the cylindrical safety factor, $v_{ti}$ is the ion thermal speed, $e$ is the elementary charge, $\ln (\Lambda)$ is the Coulomb logarithm, $\epsilon_0$ is the permittivity of free space, $\rho_i$ is the ion gryoradius, and $m_i$ is the ion mass (assumed to be deuterium), and $\mu_0$ is the permeability of free space. We  use $q_{95}$ instead of $q_{\rm cyl}$ as the fourth feature in the dimensionless feature case to capture effects of shaping (ex. triangularity) not included in the cylindrical approximation.

\subsection{Problem formulation and performance metrics} \label{subsec:prob_form}

We choose to formulate DL prediction as a supervised classification problem: we will attempt to find a model that will accurately classify plasma states as stable or in the LDL precursor phase with sufficient warning time before the instability occurs. Following standard practices, we will hold out 20\% of the discharges as the test set: these discharges will be only used to evaluate the performance of the model. The remainder of the data will be used in the training set for the models to learn on.

A discharge is classified as being in the LDL precursor phase - the ``positive'' class - for a given alarm threshold if two conditions are met: 1) the instability score rises above the alarm threshold for two or more consecutive time steps (i.e. 10ms assertion time) \textbf{and} 2) the alarm occurs $>30$ ms before the radiator destabilizes. The first condition is intended to prevent spurious alarm triggers due to an anomalous measurement at a single time step, and the second condition discounts predictions that are ``too late'' for a disruption mitigation system (DMS) to intervene. One could instead define a tardy alarm in terms of the time needed for disruption avoidance, but this would vary depending on tokamak, actuator type, and plasma scenario. For the sake of simplicity, we choose a well-defined DMS timescale for setting the late alarm threshold, and leave a more thorough treatment of disruption avoidance timescales for a later study. Specifically, we choose a minimum warning time of $30$ ms to match the time needed for actuating the ITER DMS \cite{de2016requirements}. We also note that an alarm that occurs significantly before the LDL time is still considered a true positive, as we do not want to penalize a model for providing a long warning time that could be used in practice for LDL avoidance.

In classification tasks such as this, the goal is to achieve a high true positive rate (TPR) and low false positive rate (FPR). Concretely, the TPR is the fraction of LDL shots (the ``positive'' case) that are correctly predicted to be an LDL, and the FPR is the fraction of stable shots (the ``negative'' case) that are incorrectly predicted to have an LDL event.

For any proximity-to-instability model that provides a continuous instability score, we must choose an alarm threshold above which to predict the shot will end in an LDL. For example, $n/n_G = 1$ is often considered the standard threshold for the LDL, but the threshold could be adjusted to change the sensitivity level. A lower alarm threshold will be more sensitive, and have a higher TPR and FPR. On the flip side, a higher alarm threshold will have a lower TPR and FPR. In sum, all proximity-to-instability models have a tradeoff between TPR and FPR.

We report two performance metrics for each model: ``Area Under the Curve'' (AUC) and the FPR at a fixed TPR of 95\% (shorthand: FPR @ TPR = 95\%). The AUC is the average TPR across the range of FPR $\in [0,1]$, which quantifies the performance of the model across the full range of alarm threshold levels. The FPR @ TPR = 95\% metric, by contrast, represents the proportion of stable discharges that are incorrectly classified when we require exceptional detection performance of LDLs. This is important for ITER and future tokamak power plants, where the potential damage from disruptions necessitates near-perfect (TPR $\geq 95$\%) prediction of disruptions.

\subsection{LDL prediction models} \label{subsec:models}

We hypothesize that we can achieve higher LDL prediction accuracy than the Greenwald fraction by training data-driven models to discriminate the stable and LDL precursor phase. The motivation for this choice is that the precursor phase is a distinct and relatively long-lived regime, which provides a coherent target for the models to fit. The pitfall of this approach is that it results in a subtle difference between the training objective (classifying time steps as stable or LDL precursor) and the performance metrics for judging the models (classifying discharges as stable or LDL disruptions). The precursor regime is, of course, a necessary \textit{but not sufficient} condition for an LDL disruption to occur; these models could therefore be vulnerable to false positives from discharges that enter the LDL precursor regime but do not become MHD unstable. Nevertheless, we will show that this approach results in high LDL prediction accuracy, and leave the treatment of the MHD instability phase of the LDL for a future study.

In this study, we evaluate two types of sequence-to-sequence classification architectures: non-symbolic and symbolic. The two non-symbolic models are standard machine learning workhorses - the neural network (NN) and random forest (RF). Details about hyperparameters scans are reported in \ref{sec:hyperparams}. Hyperparameters are selected via the maximum AUC on the validation set, a randomly assigned set of 20\% of discharges withheld from the training set.

We also attempt to find a symbolic density limit boundary using two methods: linear regression and linear support vector machines (LSVM). Symbolic models are simply models that take on an analytic form (for example, the Greenwald fraction is a symbolic model). To identify the linear regression boundary, we average over the last 50 ms before the LDL and use multivariate linear regression to find a power law that minimizes the mean squared error over the training set. We encourage a parsimonious model by computing the p-value of each feature, removing the feature with the largest p-value above 0.05, and re-training until all features in the regression model have p-values less than 0.05. We find a power law boundary using LSVMs by training a classifier on all data points in the training set (just as we do for the NN and RF). Feature combinations are explored via sequential feature selection with backward elimination using the Bayesian Information Criterion as the evaluation metric,
\begin{equation}
    {\rm BIC}  = -2 \ln(L) +  k \ln(s),
\end{equation}
where $L$ is the likelihood of the model evaluated on the training set, $k$ is the number of regression variables in the model, and $s$ is the number of samples. The BIC balances a reward for low error (low negative log-likelihood) with a penalty for more parameters in the model, weighted by the log of the number of samples used to fit the parameters. As plasmas states are dynamically evolving in time during discharges and not independently sampled, we approximate $s$ as the number of discharges. We similarly adjust the likelihood $L \equiv \sum y \ln (p) + (1-y) \log (1-p)$ by rescaling it by the ratio of number of discharges to number of time steps.

Finally, we will compare the model predictions with that of the Greenwald fraction using the line-average density and the edge density. These scalings will be used as baselines for comparison to the data-driven approaches.

\section{Results} \label{sec:results}

\subsection{Predicting the LDL with ``global'' features} \label{subsec:global}

Table \ref{tab:eng_results} shows the test set performance of L-mode density limit (LDL) prediction trained on the ``global'' features (Table \ref{tab:feature_list}) in comparison to the Greenwald scaling. The symbolic boundaries are all written as proportionalities as the magnitude of the coefficient adjusts the alarm threshold. We cannot compute the Greenwald fraction for the edge density as the discharges from EAST lack this signal. 

\begin{table}
\caption{The test set performance of LDL prediction for models trained on the global features for all devices: AUG, C-Mod, DIII-D, EAST, and TCV. The best performance for each metric (highest AUC, lowest FPR) are bolded for emphasis. The Greenwald fraction is the least accurate of all models tested.}
\label{tab:eng_results}
\begin{indented}
\item[]\begin{tabular}{@{}lllc}
\br
Model & Analytic boundary & AUC & \shortstack{FPR @ \\ TPR = 95\%}   \\
\mr
NN & N/A & \textbf{0.943} & 28.4\%\\
RF & N/A & \textbf{0.943} & \textbf{22.6\%} \\
LSVM & $\bar{n}^{\rm limit} \sim \frac{I_p^{0.67} }{a^{1.80}}P_{\rm in}^{0.28} $ & 0.941 & 26.8\% \\
Lin. Reg. & $\bar{n}^{\rm limit}  \sim \frac{I_p^{0.75} }{a^{2.04}}P_{\rm in}^{0.17} $ & 0.925 & 39.5\% \\
Greenwald & $\bar{n}^{\rm limit}  \sim \frac{I_p}{\pi a^2}$ & 0.894 & 46.0\%  \\

\br
\end{tabular}
\end{indented}
\end{table}

\begin{figure}[!htbp]
\centering
\begin{subfigure}{.5\textwidth}
    \centering
    \includegraphics[width=0.85\linewidth]{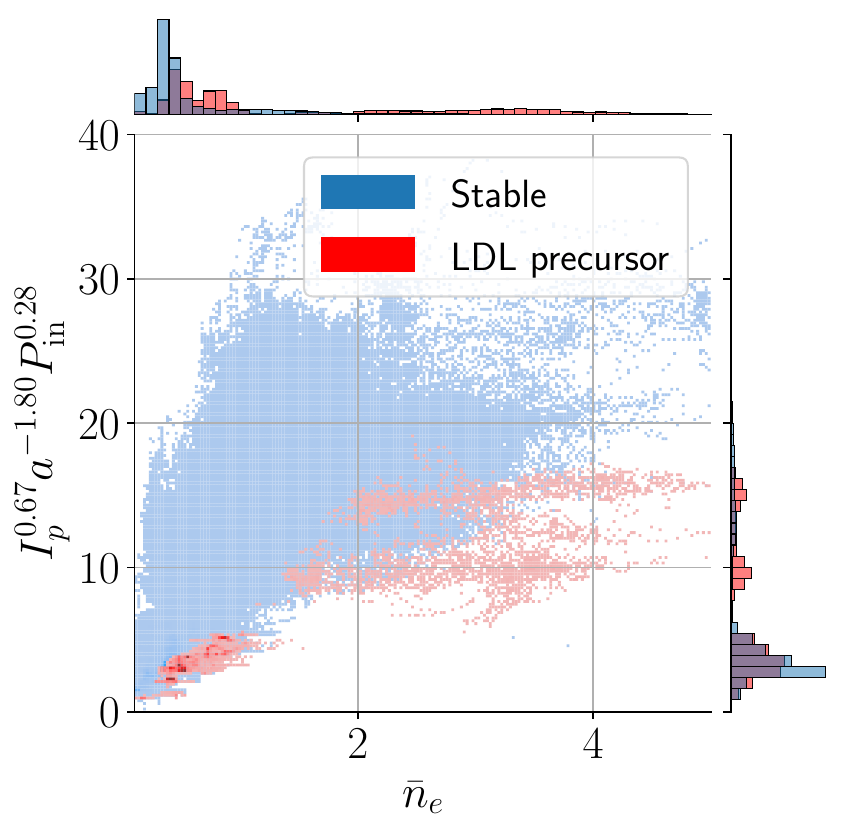}
    \caption{Labeled time steps}
    \label{subfig:eng_space_sns}
\end{subfigure}%
\begin{subfigure}{.5\textwidth}
    \centering
    \includegraphics[width=\linewidth]{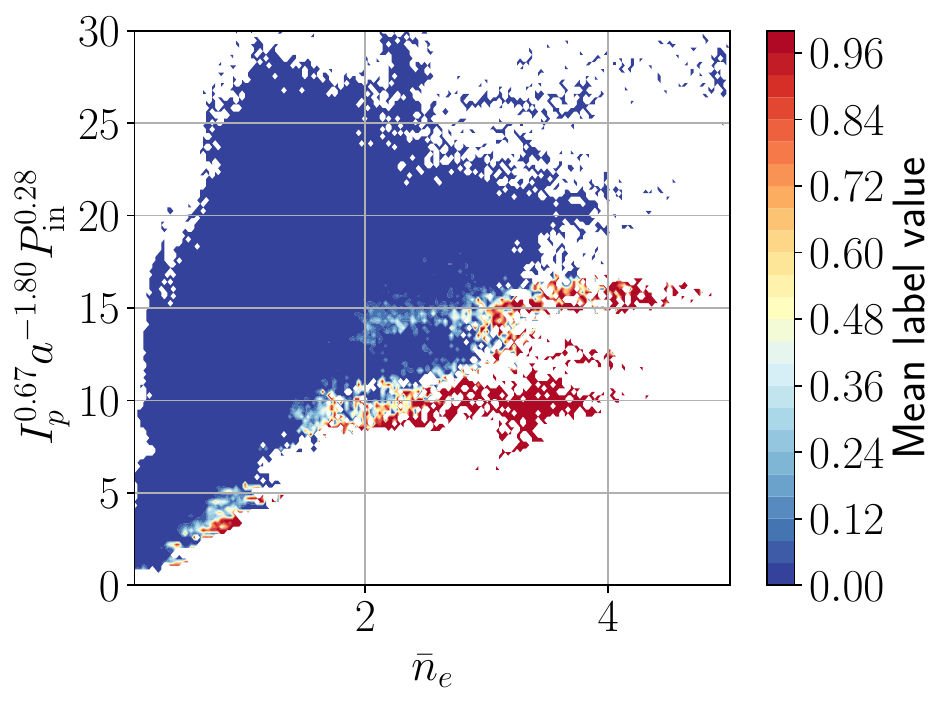}
    \caption{Mean label value }
    \label{subfig:eng_space}
\end{subfigure}
    \caption{The distribution of line averaged density vs. the remaining terms of the LSVM power law (Table \ref{tab:eng_results}) across the database. Subplot \ref{subfig:eng_space_sns} shows the LDL points (red) superimposed on the non-DL points (blue), while \ref{subfig:eng_space} shows a ``stability'' heat map, where labels (``stable'' = 0, LDL precursor = 1) have been averaged in each bin.}
    \label{fig:eng_space}
\end{figure}

We find that the NN, RF, and LSVM are the most accurate models, far outperforming the AUC and FPR of the Greenwald scaling. The linear regression model, by contrast, has performance levels between the other data-driven models and the Greenwald scaling. This illustrates the value of using a classification algorithm for this problem. Interestingly, the LSVM takes a similar form to the linear regression model -- a Greenwald-like scaling with lower current dependence and an additional $P_{\rm in }$ scaling -- but achieves nearly the same performance as the NN and RF. In Fig. \ref{fig:eng_space}, we plot the density against the remaining varliables in the LSVM power law.

Although the non-symbolic models (NN and RF) achieve higher performance than the Greenwald fraction, we note that a FPR of $>20$\% would still be very costly for the mission of ITER and FPPs. 

\subsection{Predicting the LDL with ``edge'' features} \label{subsec:dimal}

As stated previously, edge density and temperature are understood to be key parameters for the onset of the DL. Unfortunately, not all discharges in our dataset have Thomson scattering measurements of the edge. Therefore, the dataset for the ``edge'' feature analysis has a different composition of discharges, shown in Table \ref{tab:edge_num_shots}. Particularly of note is the absence of EAST data. The change in composition of the training and test set leads to different performance for the line-averaged Greenwald scaling compared to the previous section.

\begin{table}
\caption{Number of unique discharges and time steps in the database that have edge density and temperature measurements. These data are used for the analysis in section \ref{subsec:dimal} and \ref{subsec:dimless}. }
\begin{indented}
\label{tab:edge_num_shots}
\item[]\begin{tabular}{@{}lllll}
\br
Device & LDL discharges & LDL time steps & Stable discharges & Stable time steps \\
\mr
AUG & 30 & 1,063 & 8 & 15,523 \\
C-Mod & 52 & 3,322 & 2,162 & 243,221 \\
DIII-D & 41 & 2,205 & 996 & 341,717 \\
EAST & 0 & 0 & 0 & 0 \\
TCV & 30 & 1,384 & 27 & 5,776 \\
\br
Total & 153 & 7,974 & 3,193 & 606,237 \\
\end{tabular}
\end{indented}
\end{table}

As shown in Table \ref{tab:dimal_results}, the RF is the best performing model, followed closely by the NN and LSVM power law. The linear regression and Greenwald fraction scalings achieve significantly worse performance.

\begin{table}
\caption{The test set performance of LDL prediction for models trained on the edge features, as well as the Greenwald fraction and edge Greenwald fraction.}
\label{tab:dimal_results}
\begin{indented}
\item[]\begin{tabular}{@{}llll}
\br
Model & Analytic boundary & AUC & \shortstack{FPR @ \\ TPR = 95\%}   \\
\mr
NN & N/A & 0.997 & 2.8\% \\
RF & N/A & \textbf{0.998} & \textbf{0.5\%} \\
LSVM & $n_{\rm edge}^{\rm limit} \sim \frac{I_p^{0.79} }{a^{1.30}} T_{\rm edge}^{1.00}$ & 0.996 & 2.3\% \\
Lin. Reg. & $n_{\rm edge}^{\rm limit} \sim  \frac{I_p^{0.86} }{a^{1.48}} T_{\rm edge}^{0.11} q_{95}^{0.66}   $ & 0.880 & 54.6\% \\
Greenwald & $\bar{n}^{\rm limit} \sim \frac{I_p}{\pi a^2}$ & 0.971 & 13.9\%  \\
Edge Greenwald & $n_{\rm edge}^{\rm limit} \sim \frac{I_p}{\pi a^2}$ & 0.888 & 43.7\%  \\
\br
\end{tabular}
\end{indented}
\end{table}

\begin{figure}
\centering
\begin{subfigure}{.5\textwidth}
    \centering
    \includegraphics[width=0.85\linewidth]{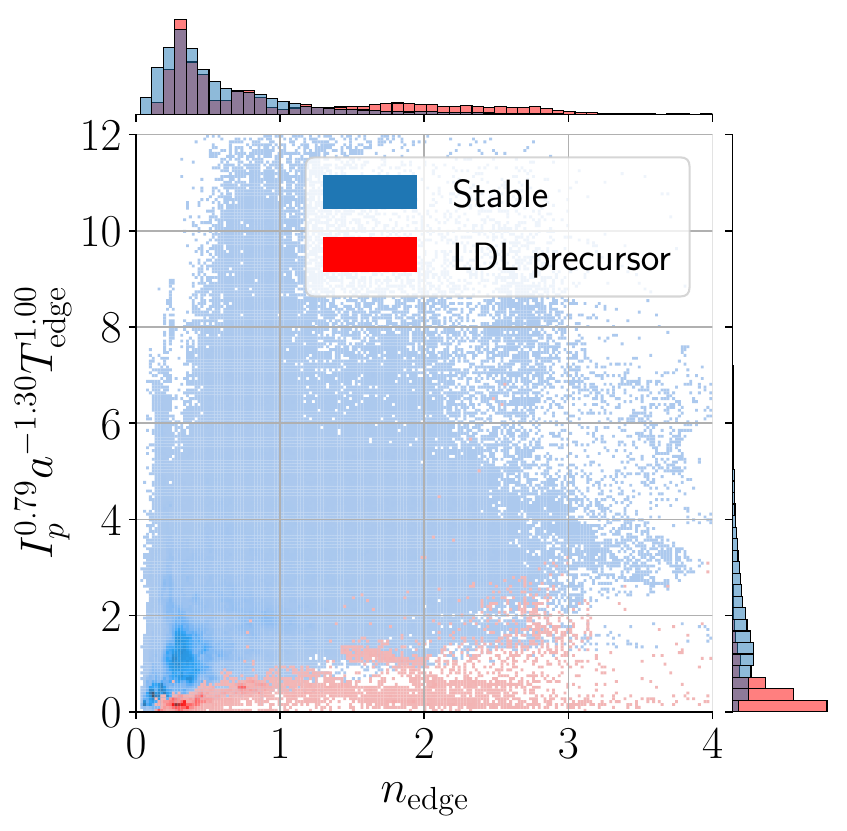}
    \caption{Labeled time steps}
    \label{subfig:dimal_space_sns}
\end{subfigure}%
\begin{subfigure}{.5\textwidth}
    \centering
    \includegraphics[width=\linewidth]{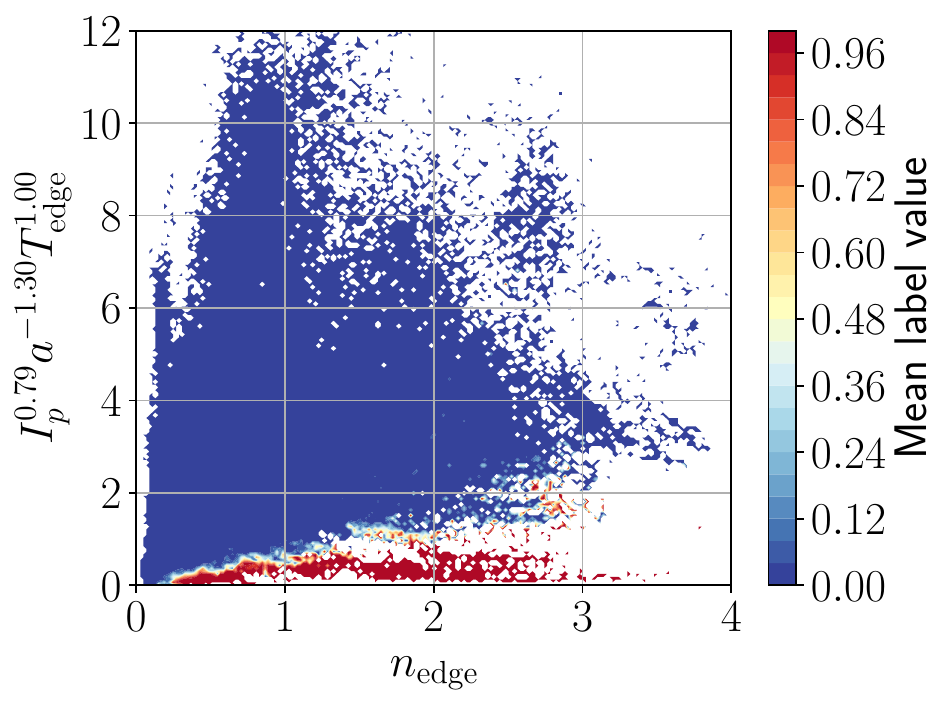}
    \caption{Mean label value }
    \label{subfig:dimal_space}
\end{subfigure}
    \caption{The distribution of edge density vs. the remaining terms of the LSVM power law for the database (Table \ref{tab:dimal_results}). Subplot \ref{subfig:dimal_space_sns} shows the LDL points (red) superimposed on the non-DL points (blue), while \ref{subfig:dimal_space} shows a ``stability'' heat map, where labels (``stable'' = 0, LDL precursor = 1)  have been averaged in each bin.}
    \label{fig:dimal_space}
\end{figure}

Interestingly, the LSVM boundary takes a similar form to the Greenwald fraction with an approximately linear temperature dependence. Calibrated to TPR = 95\%, the limit is
\begin{equation} \label{eq:dim_limit}
    n_{\rm edge}^{\rm limit} = 1.4 \frac{I_p^{0.79} }{a^{1.30}} T_{\rm edge}^{1.00}, 
\end{equation}
We show the state space of the edge density vs. the remaining terms in the LSVM power law in Fig. \ref{fig:dimal_space}. We see stronger separation of the stable and LDL precursor phases compared to the global features case in Fig. \ref{fig:eng_space}.

Once again, the linear regression power law performs significantly worse compared to the LSVM power law. The form of the boundaries are similar except for a much lower edge temperature exponent for the linear regression and the addition of a moderate $q_{95}$ dependence. The significantly diminished performance is primarily due to this the weaker temperature scaling.

We note that the Greenwald fraction model has higher performance compared to the results in section \ref{subsec:global} due to the different make-up of the dataset, as stated earlier. Nevertheless, this improved performance is still far below that of the LSVM, NN, and RF. 

\subsection{Predicting the LDL with dimensionless features} \label{subsec:dimless}

When trained on the dimensionless set of features $\nu_{*,\rm edge}$, $\rho_{*,\rm edge}$, $\beta_{T,\rm edge}$, and $q_{95}$, the data-driven models achieve similarly strong performance as is found in the edge features case (section \ref{subsec:dimal}). The test set performance metrics are reported in Table \ref{tab:dimless_results}.

\begin{table}
\caption{The test set performance of LDL prediction for models trained on the dimensionless features, as well as the Greenwald fraction and Edge Greenwald fraction.}
\label{tab:dimless_results}
\begin{indented}
\item[]\begin{tabular}{@{}llll}
\br
Model & Analytic boundary & AUC & \shortstack{FPR @ \\ TPR = 95\%}   \\
\mr
NN & N/A & 0.991 & 3.0\%\\
RF & N/A & 0.996 & \textbf{1.6\%} \\
LSVM & $\nu_{*\rm, edge}^{\rm limit}  \sim \beta_{T,\rm edge}^{-0.40}$ & \textbf{0.997} & 2.3\% \\
Lin. Reg. & $\nu_{*\rm, edge}^{\rm limit}  \sim   \beta_{T,\rm edge}^{-0.67} \rho_{*,\rm edge}^{-0.77}   $ & 0.984 & 6.6\% \\
Greenwald & $\bar{n}^{\rm limit} \sim \frac{I_p}{\pi a^2}$ & 0.971 & 13.9\%  \\
Edge Greenwald & $n_{\rm edge}^{\rm limit} \sim \frac{I_p}{\pi a^2}$ & 0.888 & 43.7\%  \\
\br
\end{tabular}
\end{indented}
\end{table}

\begin{figure}
\centering
\begin{subfigure}{.5\textwidth}
    \centering
    \includegraphics[width=0.85\linewidth]{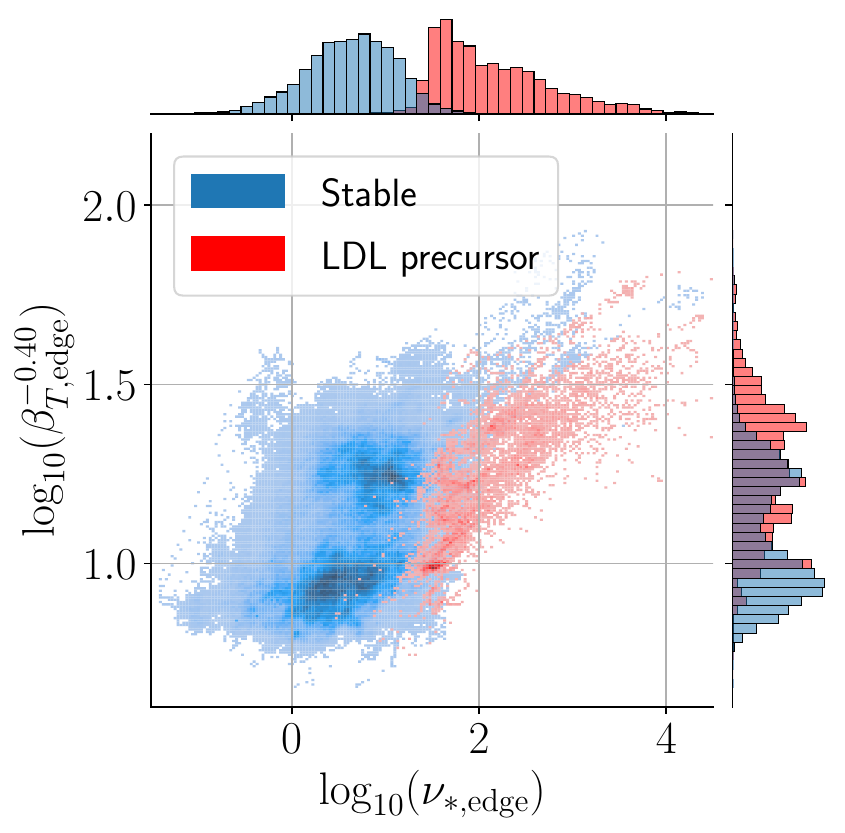}
    \caption{Labeled time steps}
    \label{subfig:dimless_space_sns}
\end{subfigure}%
\begin{subfigure}{.5\textwidth}
    \centering
    \includegraphics[width=\linewidth]{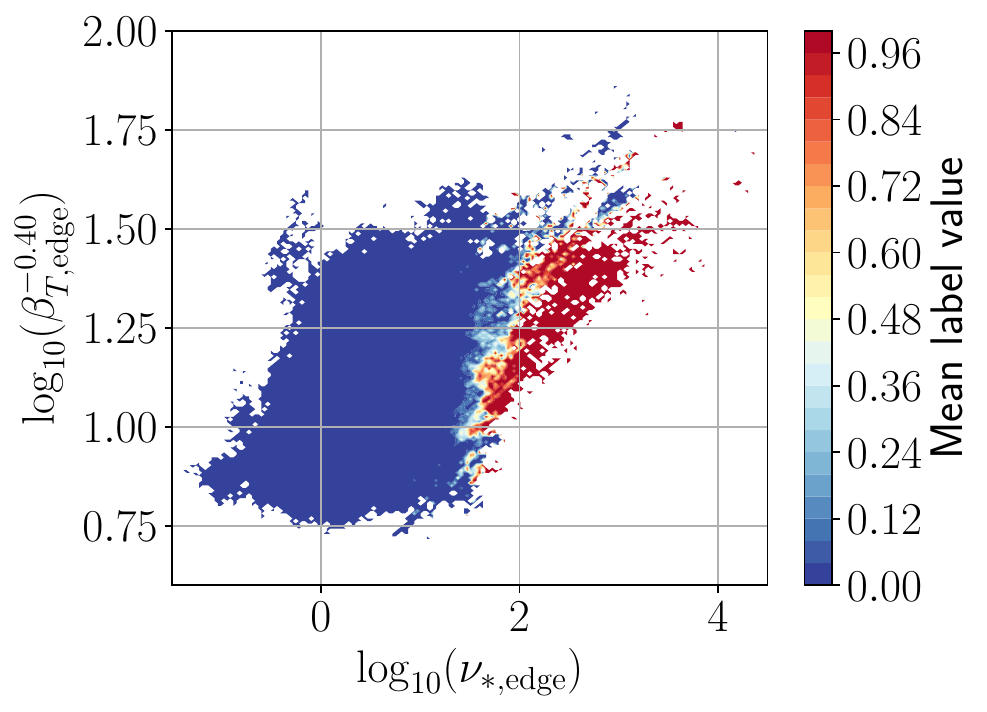}
    \caption{Mean label value }
    \label{subfig:dimless_space}
\end{subfigure}
    \caption{The distribution of edge collisionality versus $\beta_{T,\rm edge}^{-0.40}$ (Table \ref{tab:dimless_results}). Subplot \ref{subfig:dimless_space_sns} shows the LDL points (red) superimposed on the non-DL points (blue), while \ref{subfig:dimless_space} shows a ``stability'' heat map, where labels (``stable'' = 0, LDL precursor = 1) have been averaged in each bin.}
    \label{fig:dimless_space}
\end{figure}

The NN, RF, and LSVM all achieve similar AUC as in the previous section (subsection \ref{subsec:dimal}) and slightly higher FPRs. The power law boundary identified by the LSVM calibrated to TPR = 95\% is
\begin{equation} \label{eq:dimless_boundary}
    \nu_{*\rm, edge}^{\rm limit}  = 3.5\beta_{T,\rm edge}^{-0.40} .
\end{equation}
The space defined by these two variables is shown in Fig. \ref{fig:dimless_space}, illustrating relatively strong discrimination between the stable and LDL precursor points. The top marginal plot of Fig. \ref{subfig:dimless_space_sns} shows a histogram of data with respect to $\nu_{*,\rm edge}$, highlighting that the collisionality term alone provides a good degree of discrimination between the cases. 

\section{Discussion} \label{sec:discuss}

\begin{table}
\caption{LSVM-derived instability metrics for the LDL-precursor phase, as well as the shorthand used to refer to them.}
\label{tab:LSVM_boundaries}
\begin{indented}
\item[]\begin{tabular}{@{}lll}
\br
Feature set & Analytic boundary & Instability metric shorthand  \\
\mr
Global features & $\bar{n}^{\rm limit} \sim \frac{I_p^{0.67} }{a^{1.80}}P_{\rm in}^{0.28} $& LSVM-G \\
Edge features & $n_{\rm edge}^{\rm limit} \sim \frac{I_p^{0.79} }{a^{1.30}} T_{\rm edge}^{1.00}$  & LSVM-E  \\
Dimensionless features & $\nu_{*\rm, edge}^{\rm limit}  \sim \beta_{T,\rm edge}^{-0.40}$ & LSVM-D   \\
\br
\end{tabular}
\end{indented}
\end{table}

\subsection{Relation of results to the Greenwald limit}

Each of the LSVM-derived instability metrics bears some explicit or implicit resemblance to the Greenwald limit. For ease of reference, we assemble the LSVM boundaries in Table \ref{tab:LSVM_boundaries} and introduce shorthands for each case. 

In the case of the global feature set, the LSVM-G metric takes the form of a Greenwald-like scaling with an additional power dependence. The plasma current and input power dependencies, $I_p^{0.67}$ and $P^{0.28}$, are within the $I_p^{0.5-1.0}$ and $P^{0.2-0.6}$ ranges reported in the literature \cite{zanca2019power,giacomin2022first,manz2023power,stabler1992density,rapp1999density,huber2013impact,duesing1986first}. Additionally, the minor radius dependence ($a^{-1.80}$) is close to that of the Greenwald fraction ($a^{-2}$). These differences are somewhat subtle, but result in the LSVM-G metric achieving two times lower FPR @ TPR = 95\% and significantly higher AUC compared to the Greenwald fraction.

The LSVM-E metric is a Greenwald-like scaling in terms of edge density, with a sub-linear plasma current scaling ($I_p^{0.79}$), a smaller minor radius dependence ($a^{-1.30}$), and a linear edge temperature dependence. As with the LSVM-G metric, these differences result in far better performance than the Greenwald fraction; the LSVM-E metric achieves nearly six times lower FPR @ TPR = 95\%.

Interestingly, it can be shown the LSVM-E and LSVM-D scalings are nearly equivalent. In the dimensionless case,
\begin{equation} \label{eq:dimless_dimal_equiv1}
    \nu_{*,\rm edge}^{\rm limit} \sim \beta_{T,\rm edge}^{-0.4},
\end{equation}
can be rewritten as
\begin{equation} \label{eq:dimless_dimal_equiv2}
    n_{\rm edge}^{\rm limit} \sim \frac{I_p^{0.7} T_{\rm edge}^{1.1} }{a^{1.4}}   \Big (\ln(\Lambda)^{0.7} B_T^{0.1} \epsilon^{-0.6} \kappa^{0.7} \Big),
\end{equation}
which almost exactly matches LSVM-E. The term within the parentheses, $B_T^{0.1} \epsilon^{0.4} \kappa^{0.7}$, varies weakly across the entire database (standard deviation $<$ 10\% of the mean). Despite the fact the dimensionless case has a more restrictive set of features, the LSVM arrives at a nearly identical solution. 

\begin{figure}
    \centering
    \includegraphics[width=\linewidth]{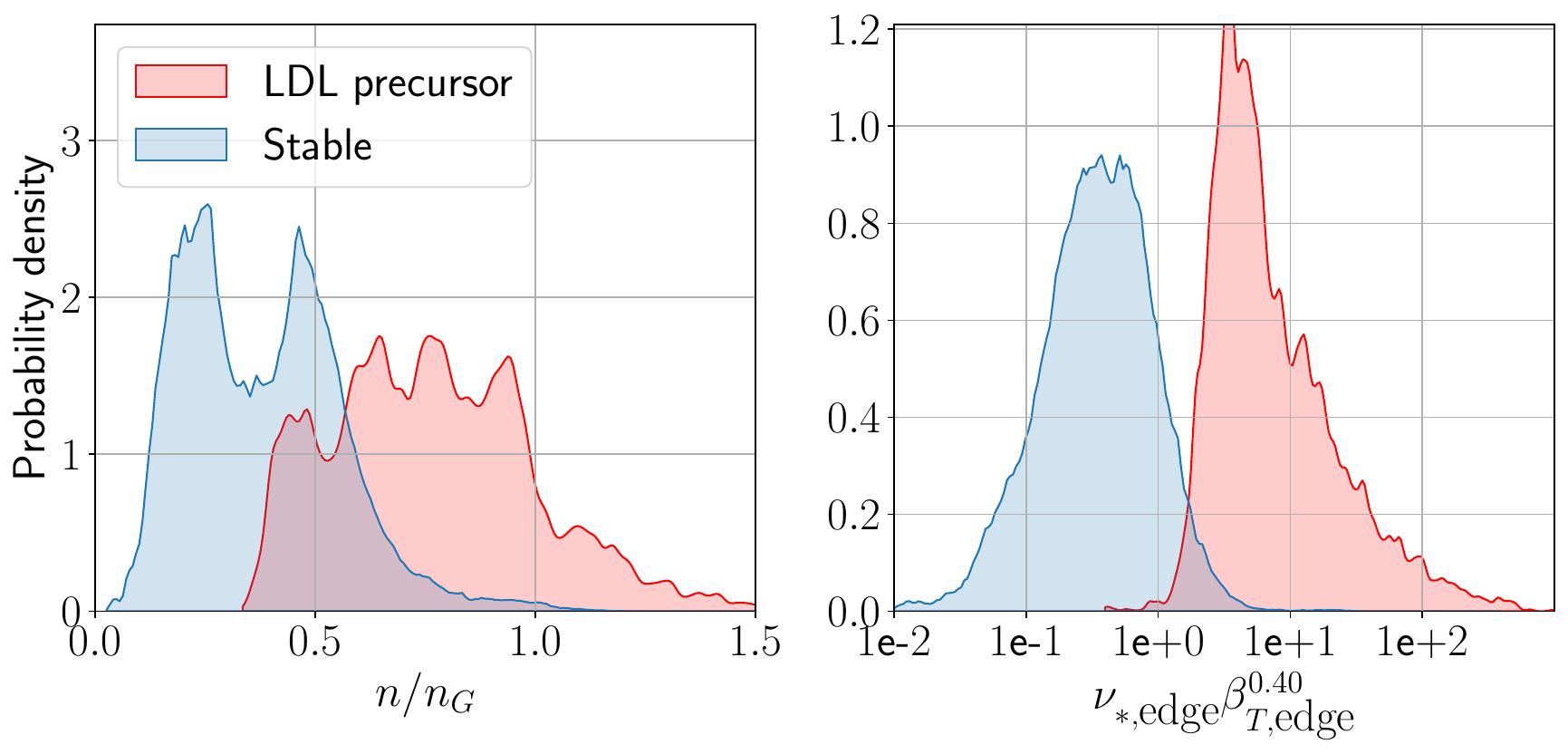}
    \caption{A comparison of the time slices in the stable and LDL precursor phase in terms of both the Greenwald fraction and LSVM-D metric. The LSVM-D metric more clearly separates the precursor phase from stable plasma states.}
    \label{fig:kde_comparison}
\end{figure}

Figure \ref{fig:kde_comparison} shows a plot of the timeslices associated with stable plasmas and the LDL precursor phase for the Greenwald fraction and LSVM-D. We see the LSVM-D metric is able to better discriminate between the LDL precursor phase and stable plasma states. Even when the LSVM is applied to a subset of machines (\ref{sec:unseen_device} and \ref{sec:other_scalings}), it finds the same pattern and achieves similar performance levels.

Notably, the $n^{\rm limit} \sim P_{\rm in}^{0.28}$ and $n^{\rm limit} \sim T_{\rm edge}^{1.00}$ dependencies identified in this study echo the approximation 
\begin{equation} \label{eq:stangeby_T}
    \bar{T}_{e,{\rm sep}} \sim P_{\rm SOL}^{2/7},
\end{equation}
where $\bar{T}_{e,{\rm sep}}$ is the average electron temperature at the last closed flux surface and $P_{\rm SOL} \equiv P_{\rm in} - P_{\rm rad}$ is the power through the SOL (input power minus power radiated within the LCFS), which is valid when parallel heat conduction dominates parallel heat convection \cite{stangeby2000plasma}. Of course $T_{\rm edge}$ is not $\bar{T}_e$, and $P_{\rm in}$ is not $P_{\rm SOL}$, but one might expect strong correlations between these parameters. 

In summary, the LSVM-derived scalings are generally consistent with past observations of a Greenwald-like scaling for the density limit and a moderate power dependence. Despite these similarities, the LSVM scalings achieve significantly improved LDL prediction accuracy.

\subsection{Relation of collisionality boundary to electron adiabaticity} \label{subsec:theo_models}

Theoretical treatments of the density limit \cite{rogers1998phase,hajjar2018dynamics} and empirical studies on individual devices \cite{hong2017edge,long2021enhanced} have suggested electron adiabaticity
\begin{equation} \label{eq:adiab}
    \alpha \equiv \frac{k_{||}^2 v_{te}^2}{\nu_{ee} \omega},
\end{equation}
in the plasma edge is a critical parameter for the density limit, where $k_{||}$ is the wavenumber along the magnetic field line (usually taken to be the inverse of the connection length $L_c \sim q R$), $v_{te}$ is the electron thermal speed,  $\nu_{ee}$ is the electron-electron collision frequency, and $\omega$ is the peak turbulence frequency. The regime $\alpha < 1$ is thought to result in increased turbulent transport through the emergence of Resistive Ballooning Modes (RBMs) \cite{rogers1998phase} or by shear layer collapse \cite{hajjar2018dynamics}. If the enhanced transport sufficiently cools the edge, the current channel narrows and an X-point radiator (XPR) or MARFE can form. Although the degradation of the shear layer is not itself a radiative mechanism, it can cause a collapse of the edge temperature to force the plasma into the radiative precursor state of the LDL.    

The adiabaticity parameter $\alpha$ is challenging to measure in practice because it involves the measuring fluctuation in the plasma edge. We can show, however, that the LSVM-D metric can be re-written in a form similar to the electron adiabaticity. Taking $k_{||} \sim 1/q_{95} R_0$ in the plasma edge (as in Ref. \cite{hong2017edge}), one can show
\begin{equation} \label{eq:dimless_adiab_equiv}
     (\nu_{*,{\rm edge}} \beta_{T,{\rm edge}}^{0.40})^{-1} \sim  \frac{ k_{||}^2 v_{te}^2 }{ \nu_{ee} \omega_{\rm imp} },
\end{equation}
where the implied frequency, $\omega_{\rm imp}$, is
\begin{equation} \label{eq:omega_imp}
    \omega_{\rm imp} \equiv \frac{ T_{\rm edge}^{0.9} }{B_T^{0.8} } \frac{n_{\rm edge}^{0.4}  k_{||} }{\epsilon^{3/2}}. 
\end{equation}
The implied frequency has temperature and magnetic field dependencies similar to those of the electron diamagnetic drift frequency
\begin{equation} \label{eq:diamag_freq}
    \omega_{*e} = \frac{T}{B}\frac{k_\perp}{en}\frac{dn}{dr}.
\end{equation}
Beyond the leading $T$ and $B$ terms, and the $\epsilon^{3/2}$ term that is relatively fixed across the database, the remaining terms do not obviously agree. We might expect a discrepancy because, as adiabiticity breaks down, the turbulence should no longer be purely drift waves.  Direct measurements of the fluctuation frequency or the density gradient at the edge would help elucidate this matter.

\subsection{Relation of collisionality boundary to Stroth et al. X-point radiator model \cite{stroth2022model}} \label{subsec:theo_model_XPR}

Reference \cite{stroth2022model} presents a scaling for the formation of an XPR by identifying the threshold at which power conducted through the edge of the plasma no longer balances the ionization and charge exchange losses. They estimate the threshold density for XPR formation to be
\begin{equation} \label{eq:stroth_cond}
n_{u}^{\rm XPR} \sim \frac{T_u^{5/2}}{n_0}\frac{a}{f_{\rm exp} R_0^2 q_s^2}
\end{equation}
where $n_{u}^{\rm XPR}$ is the upstream density for XPR formation, $T_u$ is the upstream temperature, $n_0$ is the neutral density, $f_{\rm exp}$ is the flux expansion factor, and $q_{s}$ is the safety factor of a cylindrical plasma. We cannot evaluate the full LDL model from Ref. \cite{stroth2022model} across our database, as they present a second condition involving impurity concentration for the XPR to become an unstable MARFE. However, we can arrive at an approximation of the XPR scaling 
\begin{equation} \label{eq:stroth2}
    n_{\rm edge}^{\rm limit} \sim \frac{T_{\rm edge}^{5/4} \sqrt{a} }{q_{s} R_0 }.
\end{equation}
by taking the neutral density to be proportional to the upstream density (as suggested in Ref. \cite{manz2023power}), taking the upstream density and temperature to correspond to the edge density and temperature, and assuming roughly fixed flux expansion in the XPR region across all scenarios. This expression has a similar density and temperature relationship as in LSVM-E, however we note significant discrepancies between the macroscopic parameters such as $q$ and $R_0$. Naively using eq. \ref{eq:stroth2} as a density limit indicator results in a prediction performance (AUC = 0.973, FPR = 19.7\% @ TPR = 95\%) significantly below the LSVM-derived boundaries, and similar to that of the Greenwald fraction. 

\subsection{Example discharges}

Here, we consider two example discharges to show how the Greenwald fraction and LSVM-D metric compare as LDL warning indicators. The LSVM-D metric has been calibrated for TPR = 95\%.

\begin{figure}
    \centering
    \includegraphics[width=\linewidth]{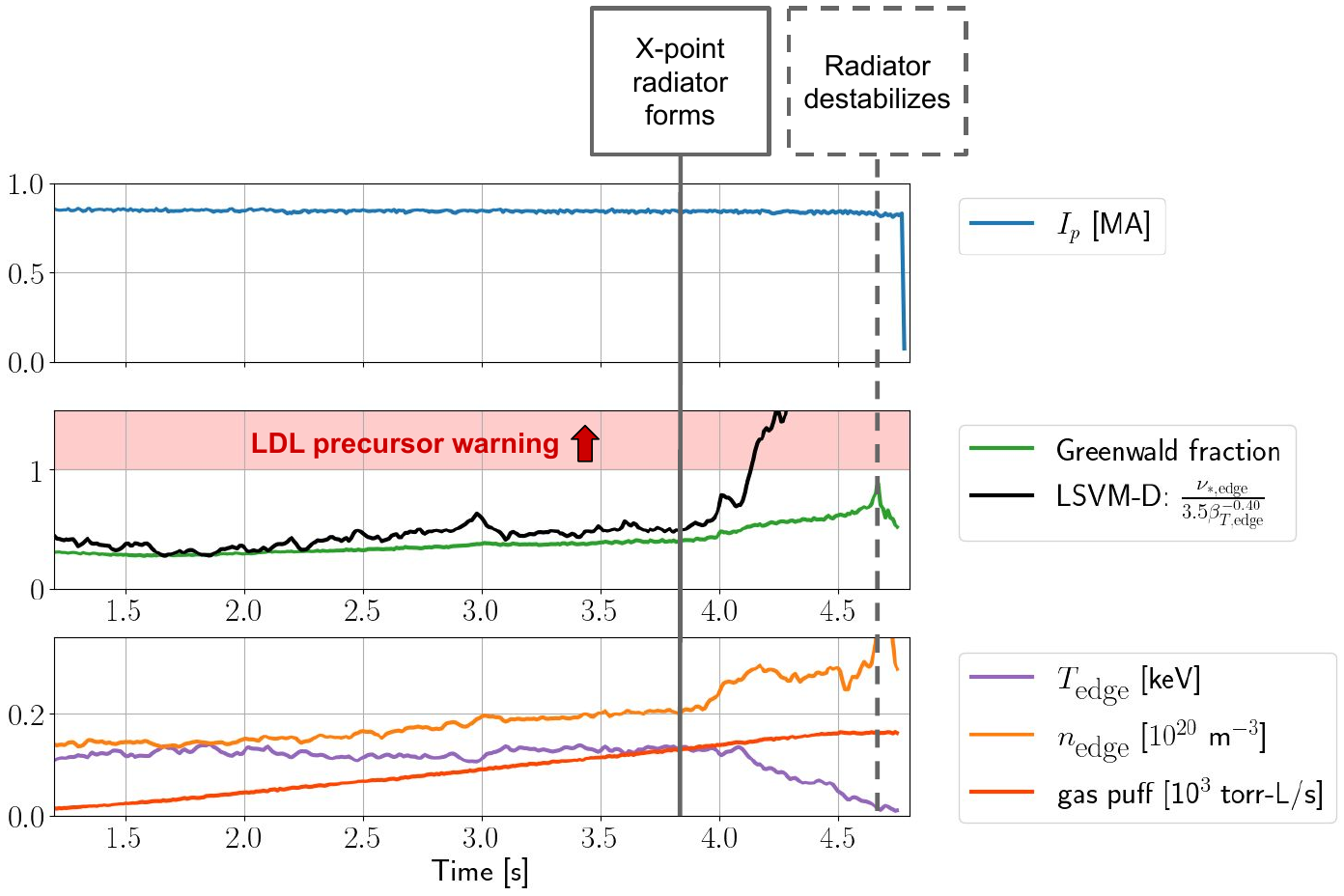}
    \caption{Time traces of DIII-D \#191794, previously shown in Fig. \ref{fig:label}, which ends in a major disruption. The middle panel shows the Greenwald fraction and LSVM-D metric calibrated to TPR = 95\%.}
    \label{fig:191794_prediction}
\end{figure}

Figure \ref{fig:191794_prediction} shows a standard density limit discharge at DIII-D (previously illustrated in Fig. \ref{fig:label} to describe the labeling). As is typical for this device, the LDL occurs at a Greenwald fraction less than 1. This would therefore be a false negative if an LDL warning threshold of $n/n_G=1$ was used. By contrast, the LSVM-G correctly warns of the LDL, rising above unity about half a second before the radiator destabilizes. This long warning time would be useful for disruption avoidance, as it would provide the control system time to recover the discharge by reducing fueling or increasing heating power. 

\begin{figure}
    \centering
    \includegraphics[width=\linewidth]{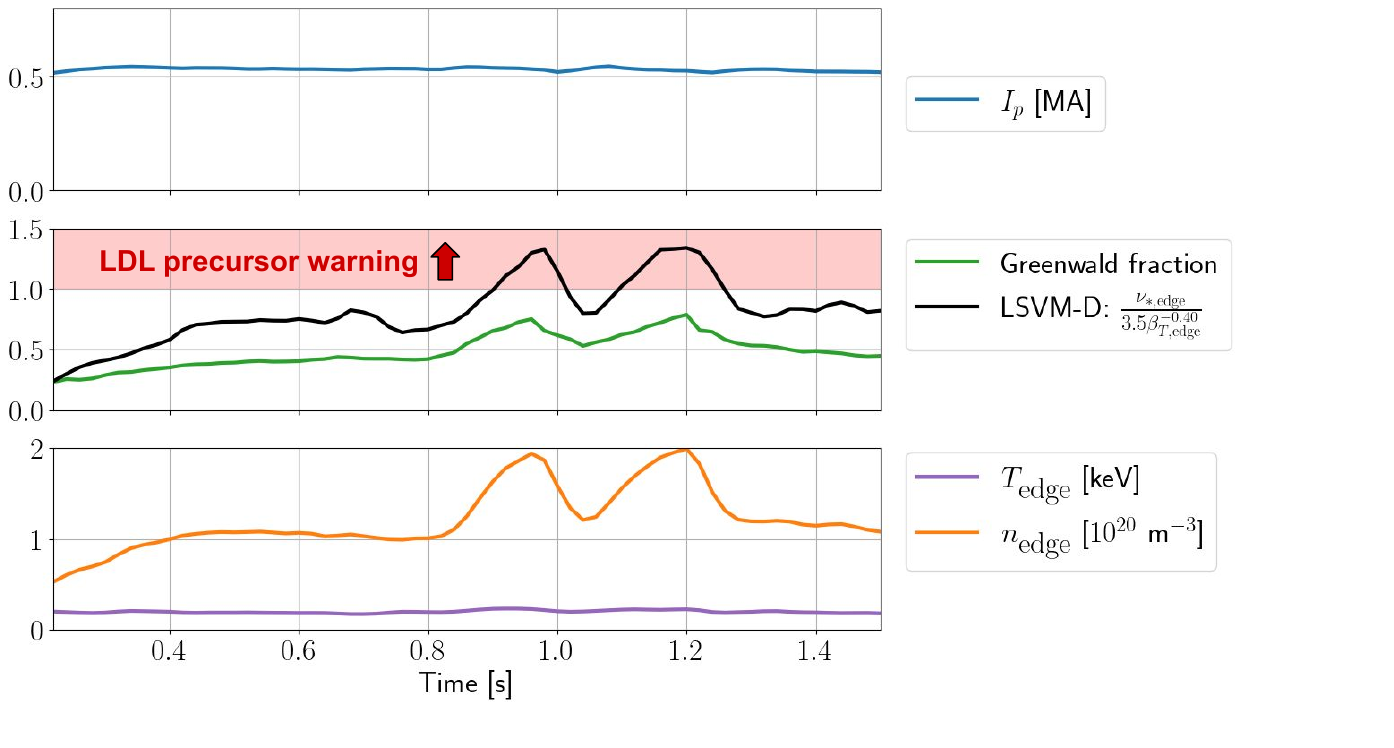}
    \caption{Time traces of non-disruptive C-Mod \#1150806028, a low-auxiliary power discharge with two brief H-modes visible in the peaks in edge density around $t=0.9$s and $t=1.2$s.}
    \label{fig:1150806028_prediction}
\end{figure}

A failure case for the LSVM-D metric is shown in Fig. \ref{fig:1150806028_prediction}. This example is the most common failure case for Alcator C-Mod: transient H-modes with low heating power. This incorrect classification may be due to the presence of the H-mode pedestal changing the correlation between the ``edge'' density and temperature (as defined in this study) with the separatrix density and temperature. If indeed the separatrix conditions set the density limit, once might expect failures when this correlation is broken.

From another perspective, this might not necessarily be a considered a failure at instability prediction, as the brief H-modes are not stable; while an LDL instability does not occur, H-to-L back-transition instabilities occur during both the excursions above the stability boundary, restoring the plasma state below the threshold after each return to L-mode. In general, other false positives for the LSVM-D can also occur for discharges with low heating power, and discharges with non-disruptive MARFEs or X-point radiators. 

\subsection{Comparison of data-driven models}

In each of the three LDL prediction cases (global features, edge features, and dimensionless features), the LSVM achieves comparable LDL prediction performance with the NN and RF. The RF has the lowest FPR in each case, but the AUC is very similar to the LSVM. This demonstrates that a highly-parameterized ML architecture such as a NN or RF is not necessary for achieving high accuracy for predicting the LDL. NNs and RFs are well suited for problems where simple, analytic functions cannot describe the observed behavior; in this case, however, a power law appears to describe the LDL precursor boundary well.

By comparing the LSVM and linear regression results, it is also evident that the way one determines the power law is critical. The linear regression approach records notably worse performance in all cases compared to the LSVM, despite the fact that both search over the same set of features and utilize the same functional form. The large gulf between these models boils down to the fact that the LSVM is utilized as a classification algorithm that leverages information from both LDL and stable plasma states, while linear regression can only use information from LDL cases. The task at hand - distinguishing LDLs from stable plasma states - is a classification problem; there is valuable information in the plasma states that end in LDLs as well as those that do not. Simply fitting an expression to the density near the LDL is not appropriate for this task.

\subsection{Limitations}

We note that the number of discharges in our database from the five devices is not uniform, as shown in Tables \ref{tab:num_shots} and \ref{tab:edge_num_shots}. In particular, the large number of stable discharges from the C-Mod and DIII-D discharges give us good statistics for the FPR in the test set, but also means that the FPR is mostly determined by discharges from those two devices.

We also note there are strong correlations among $a$, $R_0$, and $B_T$ in our database (see \ref{sec:macro_params_corr_matrix}); it is therefore impossible to disentangle the independent causal effect of these three variables. Shaping variables $\epsilon$, $\kappa$, and $\delta$ are also not included in this analysis, as there is relatively little variance (standard deviation $ \leq 25$\% of the mean) across the dataset. Additionally, there are no negative triangularity discharges in this study.

The effect of isotope mass and impurities on the density limit are also not captured in this study. The database only contains deuterium majority density limit discharges, does not include effective charge $Z_{\rm eff}$ as a parameter, and excludes discharges where the operators noted a major impurity injection. We note that the LSVM metrics achieve high accuracy across metal- and carbon-wall devices in this study, including shots in DIII-D with impurity seeding, but we also underline that these instability metrics should be applied to relatively clean, hydrogenic discharges.  

\subsection{Potential applicability for real-time control of burning plasmas}

The LDL instability metrics identified in this study could be used for real-time disruption prediction and plasma control. The most challenging measurements needed for the LSVM-E and -D metrics are the edge density and temperature signals computed from Thomson scattering (TS). While TS systems have low sampling rates relative to other diagnostics, such as the magnetics, TS measurements are frequent relative to the energy and particle confinement times that set the evolution rate of the temperature and density. For example, ITER edge TS ($r/a > 0.85$) will have a temporal resolution of 10ms and spatial resolution of 5mm, compared to several seconds of energy confinement time and a 2.8m plasma minor radius \cite{bassan2016thomson}.

Estimating the LSVM-E and -D metrics in fusion power plants (FPPs) will be more challenging as TS will likely be unavailable; the large windows for gathering scattered photons conflict with tritium breeding requirements. In this case, other diagnostics (reflectometers, interferometers, ECE) would be needed to measure or infer the edge density and temperature.

In burning plasmas, however, edge collisionality will be very low due to the tremendous self-heating. Measuring the distance to the stability boundary in real time may only be necessary during the ramp-up and ramp-down phases.

\section{Conclusion} \label{sec:conc}

In this multi-machine study, we leverage a manually labelled database of density limit disruption to identify $\nu_{*,\rm edge} \beta_{T,\rm edge}^{0.40}$ as a reliable predictor of the LDL precursor phase. This instability limit achieves excellent prediction performance (AUC = 0.997, FPR = 2.3\% @ TPR = 95\%), significantly outperforming line-averaged Greenwald and edge Greenwald scalings. We are able to uncover this scaling by training an LSVM to identify the radiative precursor phase to the LDL. Other non-symbolic data-driven approaches, such as NNs and RFs achieve similar accuracy as the LSVM power law. By contrast, standard linear regression is unable to identify a highly reliable LDL instability metric.

The LSVM power law boundaries (Table \ref{tab:LSVM_boundaries}) are reminiscent of the Greenwald limit in that they favors smaller devices with higher current. However, we achieve higher prediction accuracy by accounting for $T_{\rm edge}$ explicitly in LSVM-E or through dimensionless quantities that include edge temperature in LSVM-D. Despite the different set of features, the LSVM-E and -D metrics can be shown to be nearly identical. These scalings appears to be consistent with some theoretical models of the density limit, but additional measurements of impurities and turbulent fluctuations would be needed to confirm the association.

This study also demonstrates the utility of LSVMs for identifying stability boundaries for specific events such as the LDL. For the edge and dimensionless features case, the LSVM identifies a power law with comparable performance to the highly-parameterized NN and RF models. The LSVM power law also consistently outperforms the power law identified via linear regression. These results illustrate that for a specific instability and descriptive set of features, an LSVM can identify an accurate analytic stability boundary.

This analysis is somewhat limited by non-uniform number of discharges available across devices and correlations among some parameters (ex. $a$, $R_0$, and $B_T$). Future work will seek to address these limitations by increasing the number of non-disruptive discharges from underrepresented machines in the database (AUG, TCV), expanding the database to new devices (JET), adding data from uncommon scenarios (DIII-D negative triangularity), and potentially including devices with other shapes and aspect ratios, such as spherical tokamaks.

We also discuss the potential applicability of the collisionality boundary as an instability metric for real-time control. Current experiments, such as DIII-D, and future experiments, such as ITER, have TS systems capable of measuring edge density and temperature at sufficiently high spatial and temporal resolution in real time. FPPs will face a more constrained sensing environment that preclude TS, but other diagnostics may be able to measure or infer the relevant parameters. Additionally, the low collisionality in the edge of burning plasmas may obviate the need for density limit avoidance during the flattop. We will explore applying this indicator for real-time density limit avoidance in future work.

\ack 

The authors would like to thank A. Hubbard, J. Hughes, N. Logan, and X. Chen for providing insightful feedback on this study; A. Miller for guidance in gathering C-Mod Thomson Scattering data; and M. Tobin for providing useful code for plotting. This material is based upon work supported by the U.S. Department of Energy, Office of Science, Office of Fusion Energy Sciences, using the DIII-D National Fusion Facility, a DOE Office of Science user facility, under Awards DE-FC02-04ER54698, DE-SC0014264.

This work has been carried out within the frame-work of the EUROfusion Consortium, via the Euratom Research and Training Programme (Grant Agreement No 101052200 — EUROfusion) and funded by the Swiss State Secretariat for Education, Research, and Innovation (SERI). Views and opinions expressed are however those of the author(s) only and do not necessarily reflect those of the European Union, the European Commission, or SERI. Neither the European Union nor the European Commission nor SERI can be held responsible for them. 

This work is partially supported by the National Natural Science Foundation of China under Grant number 12005264 and the International Atomic Energy Agency under Research Contract number 26478.

Disclaimer: This report was prepared as an account of work sponsored by an agency of the United States Government. Neither the United States Government nor any agency thereof, nor any of their employees, makes any warranty, express or implied, or assumes any legal liability or responsibility for the accuracy, completeness, or usefulness of any information, apparatus, product, or process disclosed, or represents that its use would not infringe privately owned rights. Reference herein to any specific commercial product, process, or service by trade name, trademark, manufacturer, or otherwise does not necessarily constitute or imply its endorsement, recommendation, or favoring by the United States Government or any agency thereof. The views and opinions of authors expressed herein do not necessarily state or reflect those of the United States Government or any agency thereof.

\section*{References}
\bibliographystyle{unsrt}

\bibliography{references}

\appendix

\section{Range of parameters in dataset} \label{sec:macro_params_corr_matrix}

\begin{figure}
    \centering
    \includegraphics[width=\linewidth]{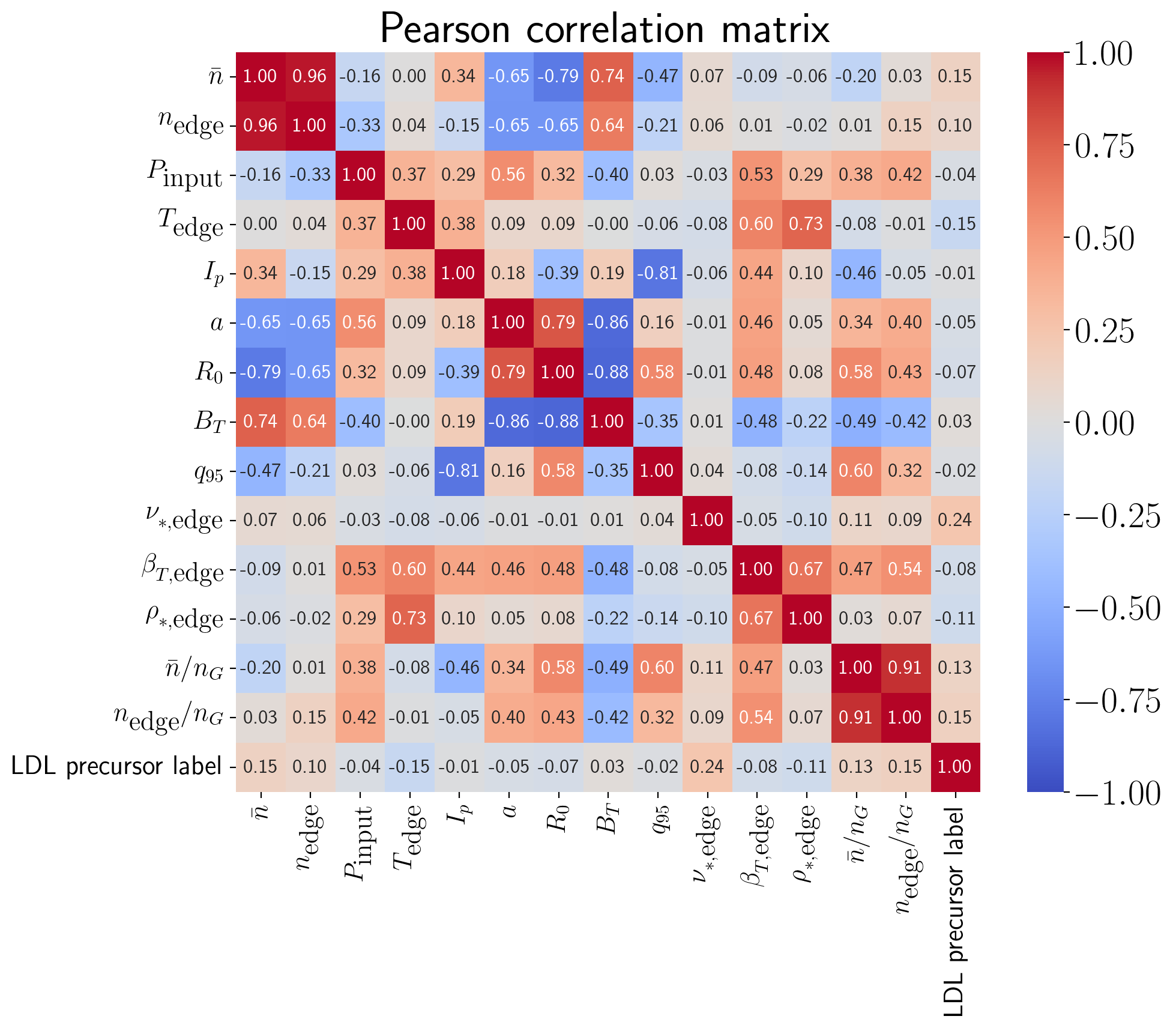}
    \caption{Correlation of several parameters in the dataset utilized in this study, including the binary LDL precursor label.}
    \label{fig:corr_matrix}
\end{figure}

Table \ref{tab:eng_sum} shows the average value and standard deviation of macroscopic parameters of the five devices in the database of this study. All parameters come from experiment measurements or equilibrium reconstructions except for the major radius of DIII-D and EAST, which are set as constant values (note: the major and minor radius of Alcator C-Mod have a finite but small standard deviation). We also show the range of dimensionless edge parameters in Table \ref{tab:dimless_sum}. The Pearson correlation matrix of several parameters used in this study are shown in Fig. \ref{fig:corr_matrix}, including the binary LDL precursor label.

\begin{table}[H]
\caption{Average value and standard deviation of macroscopic  parameters for each device in the database.}
\label{tab:eng_sum} 
\begin{indented}
\item[]\begin{tabular}{@{}l|llllll}
\br
Device & $n_e$  [$10^{20}$ m$^{-3}$] & $I_p$ [MA] & $a$ [m] & $R_0$ [m] & $B_T$ [T] & $P_{\rm in}$ [MW] \\ \hline
AUG & $0.68 \pm 0.16$ & $0.72 \pm 0.13$ & $0.50 \pm 0.01$ & $1.60 \pm 0.01$ & $2.41 \pm 0.23$ & $6.61 \pm 1.66$ \\
C-Mod & $1.45 \pm 0.68$ & $0.83 \pm 0.18$ & $0.22 \pm 0.00$ & $0.68 \pm 0.00$ & $5.44 \pm 0.81$ & $1.73 \pm 1.05$ \\
DIII-D & $0.44 \pm 0.17$ & $1.04 \pm 0.21$ & $0.59 \pm 0.02$ & $1.67 \pm 0.00$ & $1.94 \pm 0.17$ & $5.87 \pm 3.17$ \\
EAST & $0.33 \pm 0.09$ & $0.35 \pm 0.06$ & $0.44 \pm 0.01$ & $1.83 \pm 0.00$ & $2.43 \pm 0.00$ & $3.29 \pm 1.77$ \\
TCV & $0.47 \pm 0.20$ & $0.18 \pm 0.07$ & $0.23 \pm 0.01$ & $0.88 \pm 0.01$ & $1.42 \pm 0.03$ & $0.57 \pm 0.49$ \\
\end{tabular}
\end{indented}
\end{table}

\begin{table}[H]
\caption{\label{tab:dimless_sum} Average value and standard deviation of several dimensionless parameters in the edge of the plasma for each device in the database.}
\begin{indented}
\item[]\begin{tabular}{@{}l|llll}
\br
Device & $q_{95}$ & $\nu_{*\rm, edge}$ & $\beta_{T,\rm edge}$ [\%] &  $\rho^{*}_{\rm edge}$ [\%] \\ \hline 
AUG & $6.00 \pm 0.97$ & $16.51 \pm 73.24$ & $0.42 \pm 0.17$ & $0.37 \pm 0.07$ \\
C-Mod & $4.53 \pm 0.96$ & $19.97 \pm 272.04$ & $0.16 \pm 0.16$ & $0.38 \pm 0.10$ \\
DIII-D & $5.08 \pm 1.48$ & $3.69 \pm 18.74$ & $0.67 \pm 0.39$ & $0.51 \pm 0.14$ \\
EAST & $7.97 \pm 1.29$ & N/A & N/A & N/A \\
TCV & $4.85 \pm 1.19$ & $52.53 \pm 72.99$ & $0.19 \pm 0.17$ & $0.77 \pm 0.23$ \\
\end{tabular}
\end{indented}
\end{table}

\section{Hyperparameter ranges} \label{sec:hyperparams}

The neural network, random forest, and LSVM were trained over a range of hyperparameters reported in Tables \ref{tab:hyperparams_NN}, \ref{tab:hyperparams_RF}, and \ref{tab:hyperparams_LSVM}. The NN and RF hyperparameters were sampled randomly, while the LSVM hyperparameter was evaluated in a grid scan. Several sample-weighting methods were also explored, with minimal effect on the final models.

\begin{table}
\caption{\label{tab:hyperparams_NN} The hyperparameter ranges utilized for training the neural network.}
\begin{indented}
\item[]\begin{tabular}{@{}lll}
\br
Hyperparameter & Range or values & Sampling distribution  \\ \hline
Learning rate & 0.001 - 0.2 & log uniform \\
Batch size & 32, 64 & uniform \\
\# epochs & 10 - 800 & log uniform \\
\# layers & 1 - 5 & uniform \\
\# hidden units & 16, 32, 64, 128 & uniform \\
drop out proportion & 0 - 0.5 & uniform \\
activation function & relu, sigmoid & uniform \\
\end{tabular}
\end{indented}
\end{table}

\begin{table}
\caption{\label{tab:hyperparams_RF} The hyperparameter ranges utilized for training the random forest.}
\begin{indented}
\item[]\begin{tabular}{@{}lll}
\br
Hyperparameter & Range or values & Sampling distribution  \\ \hline
\# estimators & 10, 30, 100 & uniform \\
max \# features & 3 - 8 & uniform \\
max depth & 3, 5, 8 & uniform \\
min \# samples per split & 2, 5, 10 & uniform \\
min \# samples per leaf & 1, 2, 5, 10 & uniform \\
\end{tabular}
\end{indented}
\end{table}

\begin{table}
\caption{\label{tab:hyperparams_LSVM} The hyperparameter ranges utilized for training the LSVM.}
\begin{indented}
\item[]\begin{tabular}{@{}lll}
\br
Hyperparameter & Range or values  & Sampling distribution  \\ \hline
C & 0.1, 1, 10 & uniform \\
\end{tabular}
\end{indented}
\end{table}

\section{Generalizing to an unseen device} 
\label{sec:unseen_device}

As long as fusion remains an experimental science, data-driven disruption predictors must be robust to ``domain shifts,'' i.e. differences between the training set and the cases observed during deployment.  The best disruption prediction performances reported in the literature often come from highly expressive machine learning architectures, such as NNs and RFs, which can be especially vulnerable to domain shifts. Given the potentially catastrophic consequences of disruptions during a full-power discharge on ITER \cite{lehnen_plasma_2018}, robustness to domain shifts is a critical question.

Here, we consider an example of a domain shift: training on all AUG, C-Mod, and TCV discharges and then testing on DIII-D discharges. We utilize the dimensionless variables as in section \ref{subsec:dimless}, and therefore EAST is excluded due to lack of edge density and temperature measurements. DIII-D was chosen for the test set because it has the largest current, major radius, and minor radius of devices in the database that includes edge measurements.

\begin{table}
\caption{The test set performance of LDL prediction for models trained on the edge features of AUG, C-Mod, and TCV but evaluated on DIII-D. The performance of the Greenwald fraction and Edge Greenwald fraction is also reported.}
\label{tab:d3d_results}
\begin{indented}
\item[]\begin{tabular}{@{}llll}
\br
Model & Analytic boundary & AUC & \shortstack{FPR @ \\ TPR = 95\%} \\
\mr
NN & N/A & 0.987 & 3.3\% \\
RF & N/A & \textbf{0.995} & \textbf{1.5\%} \\
LSVM & $\nu_{*\rm, edge}^{\rm limit} \sim \beta_{T,\rm edge}^{-0.41} $  & 0.992 & 1.6\% \\
Lin. Reg. & $\nu_{*\rm, edge}^{\rm limit}  \sim \beta_{T,\rm edge}^{-1.06} $ & 0.974 & 9.6\% \\
Greenwald & $\bar{n}^{\rm limit} \sim \frac{I_p}{\pi a^2}$ & 0.847 & 53.7\%  \\
Edge Greenwald & $n_{\rm edge}^{\rm limit} \sim \frac{I_p}{\pi a^2}$ & 0.705 & 82.6\%  \\

\br
\end{tabular}
\end{indented}
\end{table}

The results are shown in Table \ref{tab:d3d_results}. The RF achieved the highest performance of all, with the LSVM close behind. The LSVM instability metric here is nearly identical to the one derived when the training set included DIII-D data (Section \ref{subsec:dimless}). Despite the domain shift, all data-driven models have better performance than the Greenwald scaling.

\section{Assessing Giacomin-Ricci scaling \cite{giacomin2022first} as a disruption predictor on AUG, DIII-D, and TCV} \label{sec:other_scalings}

Several theoretical models for the density limit, such as in Ref. \cite{giacomin2022first}, offer compelling explanations of the density limit.  This first-principles scaling is not explicitly intended for providing a warning to the density limit, but it could be used for this purpose. We therefore evaluate the scaling for the maximum density from Ref. \cite{giacomin2022first} to see how it fairs as a LDL predictor.

To do so, we must rely on only AUG, DIII-D, and TCV, where we have consistent measurements of the power through the SOL. Additionally, we note that the scaling in \cite{giacomin2022first} estimates the maximum density 'in the proximity of the separatrix,' not the 'edge' (as defined both here and in \cite{giacomin2022first} as the average TS measurement between $\rho = 0.85$ and $\rho = 0.95$). Just as in \cite{giacomin2022first}, however, we will proceed with utilizing the edge density in the absence of reliable measurements at the separatrix. In light of this, we emphasize that we are not attempting a complete validation of this model in this exercise. Finally, we note that our analysis overlaps in terms of AUG and TCV, but our study has data from DIII-D instead of JET and includes stable discharges to quantify the accuracy of the metric for LDL prediction.

In Table \ref{tab:giacomin_results}, we show the LDL prediction accuracy for a NN, RF, LSVM, linear regression model, the Greenwald fraction, the edge Greenwald fraction, and the Giacomin-Ricci scaling. We see that the Giacomin-Ricci scaling achieves higher AUC and lower FPR @ TPR = 95\% than the Greenwald fractions. Compared with the data-driven models, however, the Giacomin-Ricci scaling has a significantly lower performance. The LSVM boundary achieves the highest AUC and nearly matches the RF for lowest FPR @ TPR = 95\%. The LSVM boundary is nearly the same as the case presented earlier in section \ref{subsec:dimless}.

\begin{table}
\caption{ The test set performance of LDL prediction for models trained and tested on dimensionless features for AUG, TCV, and DIII-D to compare with the Giacomin-Ricci scaling \cite{giacomin2022first}. The performance of the Greenwald fraction and edge Greenwald fraction is also reported.}
\label{tab:giacomin_results}
\begin{indented}
\item[]\begin{tabular}{@{}llll}
\br
Model & Analytic boundary & AUC & \shortstack{FPR @ \\ TPR = 95\%}   \\
\mr
NN & N/A & 0.984 & 6.4\%\\
RF & N/A & 0.985 & \textbf{3.0\%} \\
LSVM & $\nu_{*\rm, edge}^{\rm limit}  \sim \beta_{T,\rm edge}^{-0.41}$ & \textbf{0.992} & 3.5\% \\
Lin. Reg. & $\nu_{*\rm, edge}^{\rm limit}  \sim \beta_{T,\rm edge}^{-1.14} $ & 0.976 & 11.4\% \\
Giacomin & $n^{\rm limit}_{\rm edge} \sim \frac{I_p^{22/21}}{a^{79/42}} \frac{P_{SOL}^{10/21} A^{1/6} R_{0}^{1/42} }{ B_T^{8/21} (1 + \kappa^2)^{1/3} }$ & 0.915 & 22.8\%  \\
Greenwald & $\bar{n}^{\rm limit} \sim \frac{I_p}{\pi a^2}$ & 0.896 & 44.1\%  \\
Edge Greenwald & $n_{\rm edge}^{\rm limit} \sim \frac{I_p}{\pi a^2}$ & 0.745 & 79.2\%  \\

\br
\end{tabular}
\end{indented}
\end{table}

Again, we emphasize that this is not an attempted validation of the Giacomin-Ricci scaling, as we do not utilize the density at the separatrix to conduct this analysis. Our focus is to analyze the scaling as a method of forecasting the LDL given readily available measurements. On this count, it improves upon the Greenwald limit, but is not as reliable as the LSVM-derived metric.

\end{document}